\documentclass[a4paper,twocolumn]{esapub}

\usepackage{natbib}
\usepackage{graphicx}
\usepackage{amssymb}
\usepackage{paralist}
\usepackage{xspace}
\usepackage{dcolumn}
\usepackage{amsmath}

\graphicspath{{/home/kreyken/mytex/figures/}}

\newcommand{\wray}{Wray\,977\xspace}
\newcommand{\gx}{GX\,301$-$2\xspace}

\newcommand{\ca}{\ensuremath{\sim}}
\newcommand{\Msun}{\ensuremath{\mbox{M}_\odot}\xspace}
\newcommand{\xte}{{\sl RXTE}\xspace}
\newcommand{\kev}{ke\kern -0.09em V\xspace}

\newcommand{\pca}{{\sl PCA}\xspace}
\newcommand{\isgri}{{\sl ISGRI}\xspace}
\newcommand{\spi}{{\sl SPI}\xspace}
\newcommand{\jemx}{{\sl JEM-X}\xspace}
\newcommand{\hexte}{{\sl HEXTE}\xspace}
\newcommand{\integral}{{\sl INTEGRAL}\xspace}
\newcommand{\err}[2]{\ensuremath{^{+#1}_{-#2}}\xspace}
\newcommand{\ecut}{\ensuremath{E_{\text{Cut}}}\xspace}
\newcommand{\efold}{\ensuremath{E_{\text{F}}}\xspace}
\newcommand{\ecyc}{\ensuremath{E_{\text{C}}}\xspace}
\newcommand{\Lsun}{\ensuremath{\mbox{L}_\odot}\xspace}

\begin{document}

\title{GX\,301$-$2 as seen by \integral}
\author[1,2]{I. Kreykenbohm}
\author[3,2]{K. Pottschmidt}
\author[3,2]{P. Kretschmar}
\author[4]{A. La Barbera}
\author[5]{L. Sidoli}
\author[6]{J. Wilms} 
\author[1]{S. Fritz}
\author[4]{A.~Santangelo}
\author[7]{W. Coburn}
\author[8]{W. A. Heindl}
\author[8]{R. E. Rothschild}
\author[1]{R. Staubert}
\affil[1]{Institut f\"ur Astronomie und Astrophysik --
  Astronomie, University of T\"ubingen, Germany}
\affil[2]{INTEGRAL Science Data Center, ISDC, 1290 Versoix,
  Switzerland} 
\affil[3]{Max-Planck-Institut f\"ur extraterrestrische Physik,
  Giessenbachstr.~1, 85740 Garching, Germany}
\affil[4]{IASF CNR Palermo, Italy}
\affil[5]{IASF Milano, Italy}
\affil[6]{Department of Physics, University of Warwick, Coventry, CV7 1AL, UK}
\affil[7]{Space Sciences Laboratory, University of California,
Berkeley, Berkeley, CA, 94702-7450, U.S.A.}
\affil[8]{CASS, University of California, San Diego, La Jolla, CA
  92093, U.S.A.}

\maketitle

\begin{abstract}
  
  We present observations of the High Mass X-ray Binary \gx taken
  during the Galactic Plane Scan (GPS) with \integral following our
  \xte observations \citep{kreykenbohm04a}. The optical companion of
  \gx is the B1Ia+ hypergiant Wray~977 with a luminosity of
  $1.3\times10^6$\,\Lsun and a mass of $\ca$48\,\Msun making the
  system one of the most massive X-ray binaries known.  The system was
  observed 24 times during the GPS thus covering many orbital phases
  including the pre-periastron flare. The source is clearly detected
  in all \integral instruments in most pointings.  The derived X-ray
  light curves show strong variability on all timescales. Depending on
  the orbital phase of the pointings, the luminosity changes by factor
  of more than 10 during the pre-periastron flare compared to other
  pointings.  Furthermore, broad band spectra using all instruments
  available on \integral are compared with data taken by the \xte
  instruments \pca and \hexte \citep[using the same data as in
  ][]{kreykenbohm04a}.
\end{abstract}

\section{\wray and \gx}

\gx consists of a 1.4\,\Msun neutron star orbiting the early
B-emission line star \wray. The neutron star is an accreting X-ray
pulsar with a rotational period of\ca685\,s and a magnetic field
strength of \ca$4\times10^{12}$\,G \citep[derived from the measurement
of a cyclotron line at \ca35\,\kev, ][]{kreykenbohm04a}.  While
\citet{parkes80a} classify \wray as a \hbox{B2 Iae} supergiant,
\cite*{kaper95a} reclassify it as a \hbox{B1 Ia+} hypergiant, derive a
distance of $d$ = 5.3\,kpc, opposed to \ca1.8\,kpc, and a higher lower
mass limit of \ca48\,\Msun.  The latter authors also estimate a
mass-loss via wind ($v_{\infty}$ = 400\,km s$^{-1}$) of
\ca10$^{-5}$\,\Msun/yr which is one of the highest wind-mass-loss
rates known. This dense wind can supply the accreting neutron star
even over a large distance (e.g., apastron) with enough material to
explain the observed luminosity of \ca10$^{37}$\,erg s$^{-1}$.

\begin{figure}
\vbox{
\centerline{\includegraphics[width=1.0\columnwidth]{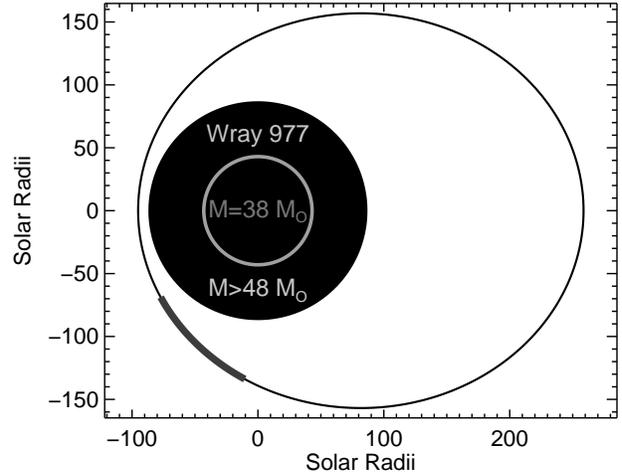}}\smallskip

\caption{Sketch of the system \gx\ / \wray based on the parameters of
  \citet{kaper95a}.  The approximate position of the pre periastron
  flare is shown in grey. The inner circle represents the size of
  \wray when using the more conservative values given by
  \citet{parkes80a}.}\vspace*{0.5mm}
\label{orbit}}
\end{figure}

\begin{figure}
\vbox{
\includegraphics[width=1.0\columnwidth]{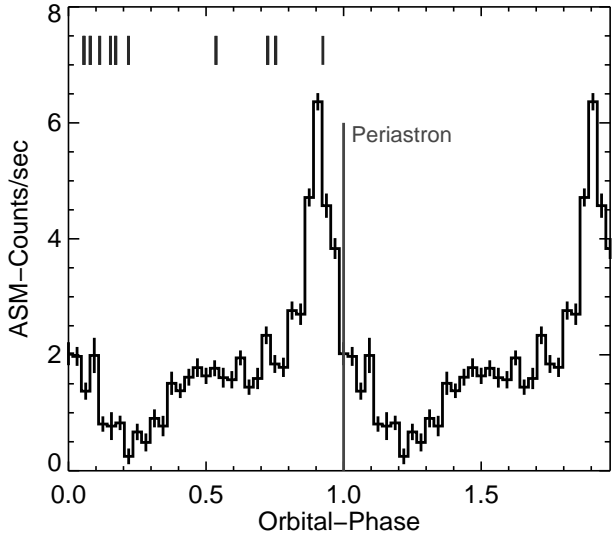}
\vspace*{1em}

\caption{Folded light curve (FLC) using all available data (starting
  in 1996 until now) on \gx of the All Sky Monitor on board \xte.
  The light curve has been folded with the orbital period of 41.498\,d
  \citep{koh97a}. The periastron passage has been extrapolated based
  on the ephemeris of \citet{koh97a}.  For clarity the folded light
  curve is shown twice. The dashes above the light curve
  indicate orbital phases during which \integral observed \gx.}
\label{asmlight}}
\end{figure}

The orbit of the neutron star has a period of 41.498$\pm$0.007\,days
\citep{koh97a} and an eccentricity of 0.472. Shortly before periastron
passage, the neutron star passes through the atmosphere at a height of
$\sim$0.1\,R$_\star$ above \wray (see Fig.~\ref{orbit}). This gives
rise to a significant fraction of the X-rays being reflected by the
stellar companion. Furthermore it also intercepts a gas stream from
the optical companion resulting in periodic extended flares which are
centered shortly before periastron passage \citep[see
Figs.~\ref{orbit},~\ref{asmlight}, and ][]{pravdo95a}.  After the
periastron passage the X-ray activity reaches a minimum to increase
slowly over the orbit with a possible second maximum at apastron
\citep{leahy91a,leahy02a}.  During the pre-periastron flare, \gx can
be 25 times brighter than in quiescence and the source is very
variable: brightness changes by a factor of two within one hour are
not uncommon \citep{rothschild87a,kreykenbohm04a}.  Following the
periastron passage is an extended low which is probably due to the
optical companion almost eclipsing the neutron star.

The spectrum of \gx is usually modeled by a power law modified at
higher energies by an exponential cutoff and photoelectric absorption
at lower energies. The spectrum is further modified by an additive
iron fluorescence line at 6.4\,\kev and a cyclotron resonant
scattering feature (CRSF) at \ca38\,\kev.

\begin{figure}
\centerline{\includegraphics[width=\columnwidth]{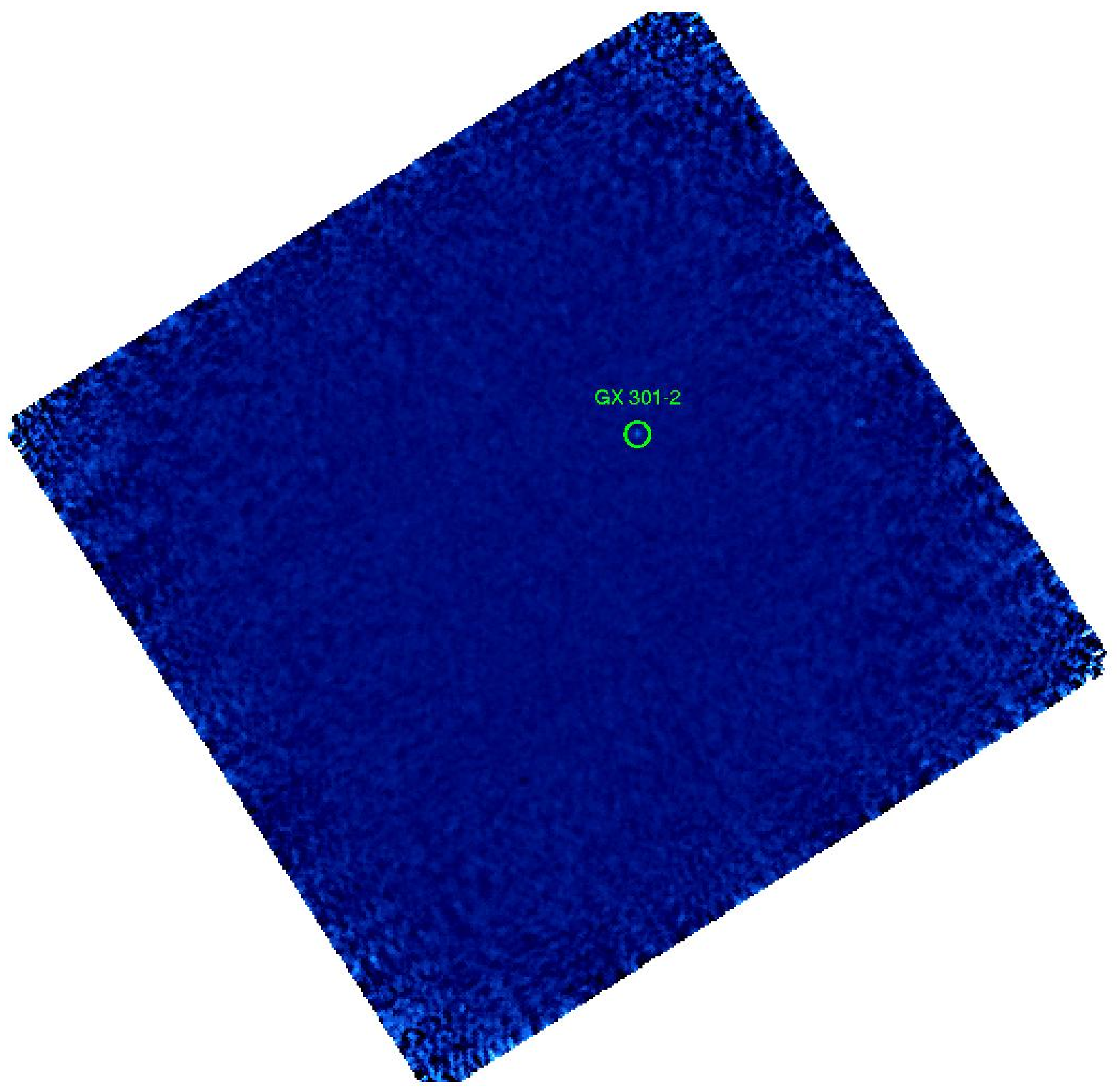}}

\caption{\isgri image (energy band 20--40\,\kev) of \gx using GPS
  data. This observation was taken during the extended low following
  the periastron passage (MJD\,52668.42). The position of \gx is
  indicated by a circle. }
\label{image_low}
\vspace*{3\baselineskip}

\centerline{\includegraphics[width=\columnwidth]{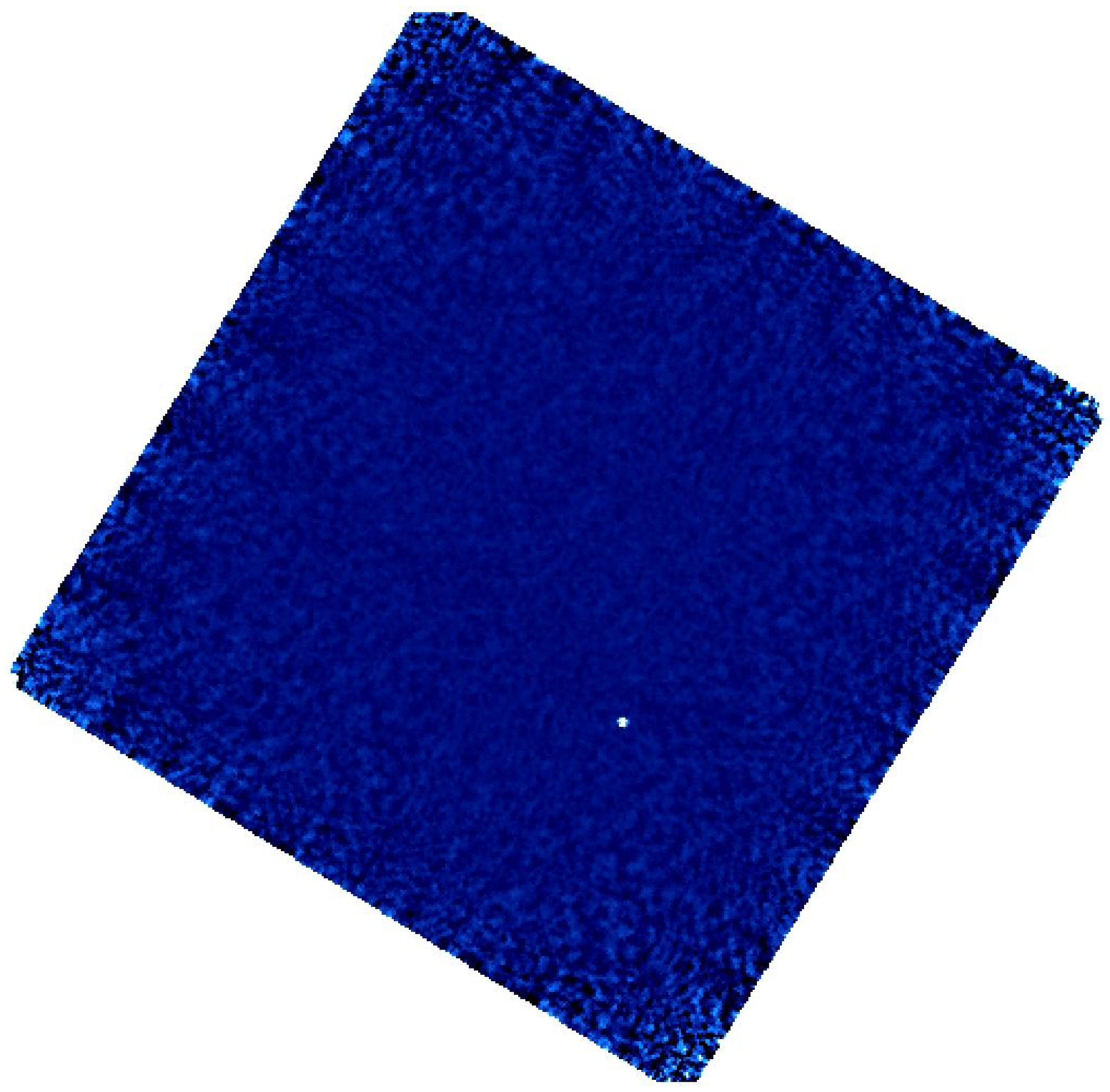}}

\caption{\isgri image (energy band 20--40\,\kev) of \gx using GPS
  data. This observation was taken during a pre-periastron flare
  (MJD\,52700.47): \gx was \ca4 times brighter than in most other ScWs
  and over 10 times brighter than during the extended low (note the
  bright spot). }
\label{image_bright}
\end{figure}

\section{Data}

\gx was observed several times by \integral during the Galactic Plane
Scan (GPS). We obtained 24 Science Windows (ScWs) during which \gx was
in the field of view (FOV) of \isgri and \spi. Since the FOV of \jemx
is considerably smaller, the source was only visible in 2 ScWs. The
observations cover almost all orbital phases, but are mostly
concentrated around orbital phase \ca0.1 (see Fig.~\ref{asmlight}),
where the source is relatively dim. In the ScWs taken when \gx is in
the extended low around orbital phase 0.2, the source is only
marginally detected. Fortunately, we also have one observation during
the pre-periastron flare. (see Fig.~\ref{image_low} and
Fig.~\ref{image_bright}).  During this pre-periastron flare the
luminosity of the source increased by a factor of \ca10: compare
Fig.~\ref{image_bright} and Fig~\ref{image_low} when the source was in
the extended low.

\section{Spectra}

We derived spectra for all ScWs for \spi and \isgri using OSA version
3.0 which were then combined to obtain spectra of higher statistical
quality. To model the resulting spectra, we used several models
including a single power-law, a power-law modified by an exponential
cutoff (in \textsl{XSPEC}: \texttt{cutoffpl}), and a power-law
modified at higher energies by the Fermi-Dirac cutoff
\citep[FDCO,][]{tanaka86a}:

\begin{equation}
I_{\textsf{NS(E)}}(E) = A_{\textsf{PL}}
\frac{E^{-\Gamma}}{\exp\left((E-\ecut)/\efold\right)+1} 
\end{equation}

The conventional high energy cutoff of \citet{white83a} cannot be used
as it produces line like residuals at the cutoff energy
\citep{kreykenbohm99a}.  While the FDCO and the cutoffpl produce
almost identical results in terms of resulting parameters (see
Table~\ref{fits}) and $\chi^2$, simpler models like a single power law
cannot describe the data at all. The resulting parameters show clearly
that low energy coverage is required: for the FDCO model, the cutoff
energy could not be determined and had to be fixed to a typical value.

\begin{figure}
\includegraphics[width=\columnwidth]{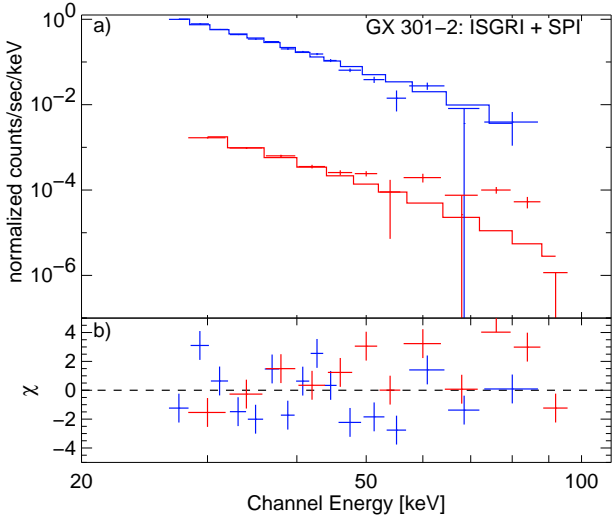}

\caption{\textbf{a} Data (exposure time \ca55\,ksec) and folded model
  of the spectrum of \gx (top: \isgri, bottom: \spi).  \textbf{b}
  Residuals of the fit.}
\end{figure}

To get coverage at energies below \ca25\,\kev, we used a 200\,ksec
\xte observation taken in 2000 covering most of the pre-periastron
flare and the periastron passage \citep{kreykenbohm04a}.  Since a
power law plus cutoff cannot describe the spectrum of \gx below
\ca20\,\kev \citep{kreykenbohm02a}, we used an absorbed partial
covering model \citep[PC; as described by ][]{kreykenbohm04a} to
simultaneously fit \integral and \xte data (see Table~\ref{fits}).
The high values for the high as well as the less absorbed component
show that the neutron star is indeed deeply embedded in the dense
stellar wind of the optical companion.

\begin{figure}
\includegraphics[width=\columnwidth]{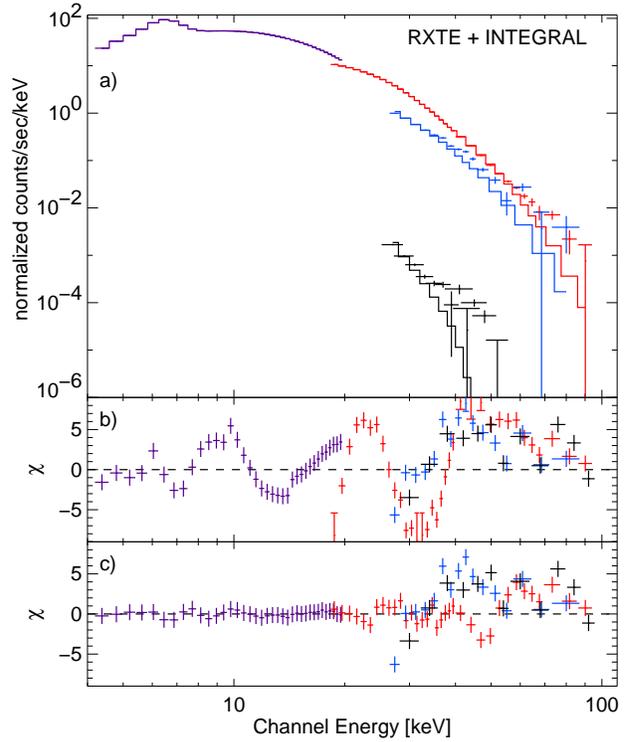}

\caption{\textbf{a} Data and folded model (partial covering model plus
  a CRSF; see text for details) of \gx using instruments of \xte
  (\pca, 3\,\kev--20\,\kev, and right, top: \hexte, 20--80\,\kev) and
  \integral (right, middle: \isgri and right, bottom: \spi).
  \textbf{b} residuals for fit without a CRSF and \textbf{c}
  residuals when including a CRSF at 34.0$\pm0.5$\,\kev with
  a width of 4.2$\pm0.5$\,\kev and a depth $\tau=0.14\pm0.01$.  }
\label{joint}
\end{figure}

At higher energies, the \hexte and the \integral instruments do not
agree very well, as shown in Fig.~\ref{joint}. While the \hexte
clearly shows the presence of a cyclotron feature at \ca35\,\kev,
there is only an indication for a feature at the same energy in the
\isgri data. Due to remaining calibration issues we are not able at
this moment to confirm the cyclotron scattering feature with the
\isgri data. To clearly detect the CRSF in \spi data, significantly
more data are required.

\begin{table}
\caption{Fit results. The models in use are a power law $+$
  Fermi-Dirac cutoff (FDCO), the cutoffpl, and a partial covering
  model (PC). }\medskip
\label{fits}
\begin{tabular}{lrrr}
\hline
\hline
 & Integral & Integral & Integral \\
&  & & + \xte \\
& FDCO & cutoffpl & PC \\
\hline
$N_{\text{H},1}$ & --- & --- & 28.9\err{0.3}{0.4} \\
$N_{\text{H},2}$ & --- & --- & 238\err{15}{12} \\
\ecut & 20 fix & ---  & 19.7\err{0.5}{3.0} \\
\efold & 21.6\err{1.7}{1.8} & 20.5\err{1.2}{1.7} & 5.7\err{0.1}{0.1} \\
$\Gamma$ & 3.7\err{0.7}{0.4} & 3.1\err{1.0}{0.6} & 0.4\err{0.1}{0.1} \\
\ecyc & --- & --- & 33.7\err{0.6}{0.4} \\
$\sigma_\text{cyc}$ & --- & --- & 7.5\err{0.5}{0.4} \\
$\tau_\text{cyc}$ & --- & --- & 0.28\err{0.06}{0.02} \\
\hline
\end{tabular}
\end{table}

\bibliographystyle{jwaabib} \parskip0pt \bibsep0pt
\bibliography{mnemonic,velax1,div_xpuls,foreign,xpuls,cyclotron,roentgen,books}

\begin{thebibliography}{}

\bibitem[\protect\astroncite{Kaper et~al.}{1995}]{kaper95a}
Kaper L., Lamers H.J.G.L.M., Ruymaekers E., et~al., 1995, Astron. Astrophys.
  300, 446

\bibitem[\protect\astroncite{Koh et~al.}{1997}]{koh97a}
Koh D.T., Bildsten L., Chakrabarty D., et~al., 1997, Astrophys. J. 479, 933

\bibitem[\protect\astroncite{Kreykenbohm et~al.}{2001}]{kreykenbohm02a}
Kreykenbohm I., Coburn W., Wilms J., et~al., 2001,
\newblock In: Jansen F., TBD (eds.) Proc. 'New Visions of the {X}-ray Universe
  in the {XMM}-Newton and {C}handra Era'. ESA SP-488, ESTEC, The Netherlands

\bibitem[\protect\astroncite{Kreykenbohm et~al.}{1999}]{kreykenbohm99a}
Kreykenbohm I., Kretschmar P., Wilms J., et~al., 1999, Astron. Astrophys. 341,
  141

\bibitem[\protect\astroncite{Kreykenbohm et~al.}{2004}]{kreykenbohm04a}
Kreykenbohm I., Wilms J., Coburn W., et~al., 2004, Astron. Astrophys. submitted

\bibitem[\protect\astroncite{Leahy}{1991}]{leahy91a}
Leahy D.A.,  1991, Mon. Not. R. Astron. Soc. 250, 310

\bibitem[\protect\astroncite{Leahy}{2002}]{leahy02a}
Leahy D.A.,  2002, Astron. Astrophys. 391, 219

\bibitem[\protect\astroncite{Parkes et~al.}{1980}]{parkes80a}
Parkes G.E., Mason K.O., Murdin P.G., Culhane J.L.,  1980, Mon. Not. R. Astron.
  Soc. 191, 547

\bibitem[\protect\astroncite{Pravdo et~al.}{1995}]{pravdo95a}
Pravdo S.H., Day C.S.R., Angelini L., et~al., 1995, Astrophys. J. 454, 872

\bibitem[\protect\astroncite{Rothschild \& Soong}{1987}]{rothschild87a}
Rothschild R.E., Soong Y.,  1987, Astrophys. J. 315, 154

\bibitem[\protect\astroncite{Tanaka}{1986}]{tanaka86a}
Tanaka Y.,  1986,
\newblock In: Mihalas D., Winkler K.H.A. (eds.) Radiation Hydrodynamics in
  Stars and Compact Objects. IAU Coll. 89, Springer, Heidelberg, p. 198

\bibitem[\protect\astroncite{White et~al.}{1983}]{white83a}
White N.E., Swank J.H., Holt S.S.,  1983, Astrophys. J. 270, 711

\end{thebibliography}

\end{document}